\documentclass[fleqn,usenatbib]{mnras}

\usepackage{newtxtext,newtxmath}

\usepackage[T1]{fontenc}

\usepackage{subcaption}

\DeclareRobustCommand{\VAN}[3]{#2}
\let\VANthebibliography\thebibliography
\def\thebibliography{\DeclareRobustCommand{\VAN}[3]{##3}\VANthebibliography}

\usepackage{graphicx}	
\usepackage{amsmath}	

\usepackage{comment}
\specialcomment{instructions}{\begingroup\noindent\color{ForestGreen}}{\medskip\endgroup}
\excludecomment{instructions}

\title[Chromaticity of jitter in J0437$-$4715]{An insight into chromatic behaviour of jitter in pulsars and its modelling: A case study of PSR J0437$-$4715}

\author[A. D. Kulkarni et al.]{
A. D. Kulkarni,$^{1,2}$\thanks{E-mail: adkulkarni@swin.edu.au}
R. M. Shannon,$^{1,2}$
D. J. Reardon,$^{1,2}$
M. T. Miles,$^{1,2}$ 
M. Bailes,$^{1,2}$ 
M. Shamohammadi,$^{1,2}$ 
\\
$^{1}$Centre for Astrophysics and Supercomputing, Swinburne University of Technology, PO Box 218, Hawthorn, VIC 3122, Australia\\
$^{2}$ARC Centre of Excellence for Gravitational Wave Discovery (OzGrav), Mail H29, Swinburne University of Technology, PO Box
218,\\ Hawthorn, VIC 3122, Australia\\
}

\date{Accepted XXX. Received YYY; in original form ZZZ}

\pubyear{2023}

\begin{document}
\label{firstpage}
\pagerange{\pageref{firstpage}--\pageref{lastpage}}
\maketitle

\begin{abstract}
Pulse-to-pulse profile shape variations introduce correlations in pulsar times of arrival (TOAs) across radio frequency measured at the same observational epoch. This leads to a broadband noise in excess of radiometer noise, which is termed pulse jitter noise. The presence of jitter noise limits the achievable timing precision and decreases the sensitivity of pulsar-timing data sets to signals of interest such as  nanohertz-frequency gravitational waves. Current white noise models used in pulsar timing analyses attempt to account for this, assuming complete correlation of uncertainties through the arrival times collected in a unique observation and no frequency dependence of jitter (which corresponds to a rank-one covariance matrix). However, previous studies show that the brightest  millisecond pulsar at decimetre wavelengths,  PSR J0437$-$4715, shows decorrelation and frequency dependence of jitter noise. Here we present a detailed study of the decorrelation of jitter noise in PSR J0437$-$4715 and implement a new technique to model it. We show that the rate of decorrelation due to jitter can be expressed as a power-law in frequency. We analyse the covariance matrix associated with the jitter noise process and find that a higher-rank-approximation is essential to account for the decorrelation and to account for  frequency dependence of jitter noise. We show that the use of this novel method significantly improves the estimation of other chromatic noise parameters such as dispersion measure variations. However, we find no significant improvement in errors and estimation of other timing model parameters suggesting that current methods are not biased for other parameters, for this pulsar due to this misspecification.  We show that pulse energy variations show a similar decorrelation to the jitter noise, indicating a common origin for both observables.
\end{abstract}




\begin{instructions}

\section{Introduction}

This is a simple template for authors to write new MNRAS papers.
See \textsc{mnras\_sample.tex} for a more complex example, and \textsc{mnras\_guide.tex}
for a full user guide.

All papers should start with an Introduction section, which sets the work
in context, citets relevant earlier studies in the field by \citett{Fournier1901},
and describes the problem the authors aim to solve \citetp[e.g.][]{vanDijk1902}.
Multiple citations can be joined in a simple way like \citett{deLaguarde1903, delaGuarde1904}.

\section{Methods, Observations, Simulations etc.}

Normally the next section describes the techniques the authors used.
It is frequently split into subsections, such as Section~\ref{sec:maths} below.

\subsection{Maths}
\label{sec:maths} 

Simple mathematics can be inserted into the flow of the text e.g. $2\times3=6$
or $v=220$\,km\,s$^{-1}$, but more complicated expressions should be entered
as a numbered equation:

\begin{equation}
    x=\frac{-b\pm\sqrt{b^2-4ac}}{2a}.
	\label{eq:quadratic}
\end{equation}

Refer back to them as e.g. equation~(\ref{eq:quadratic}).

\subsection{Figures and tables}

Figures and tables should be placed at logical positions in the text. Don't
worry about the exact layout, which will be handled by the publishers.

Figures are referred to as e.g. Fig.~\ref{fig:example_figure}, and tables as
e.g. Table~\ref{tab:example_table}.

\begin{figure}
	\includegraphics[width=\columnwidth]{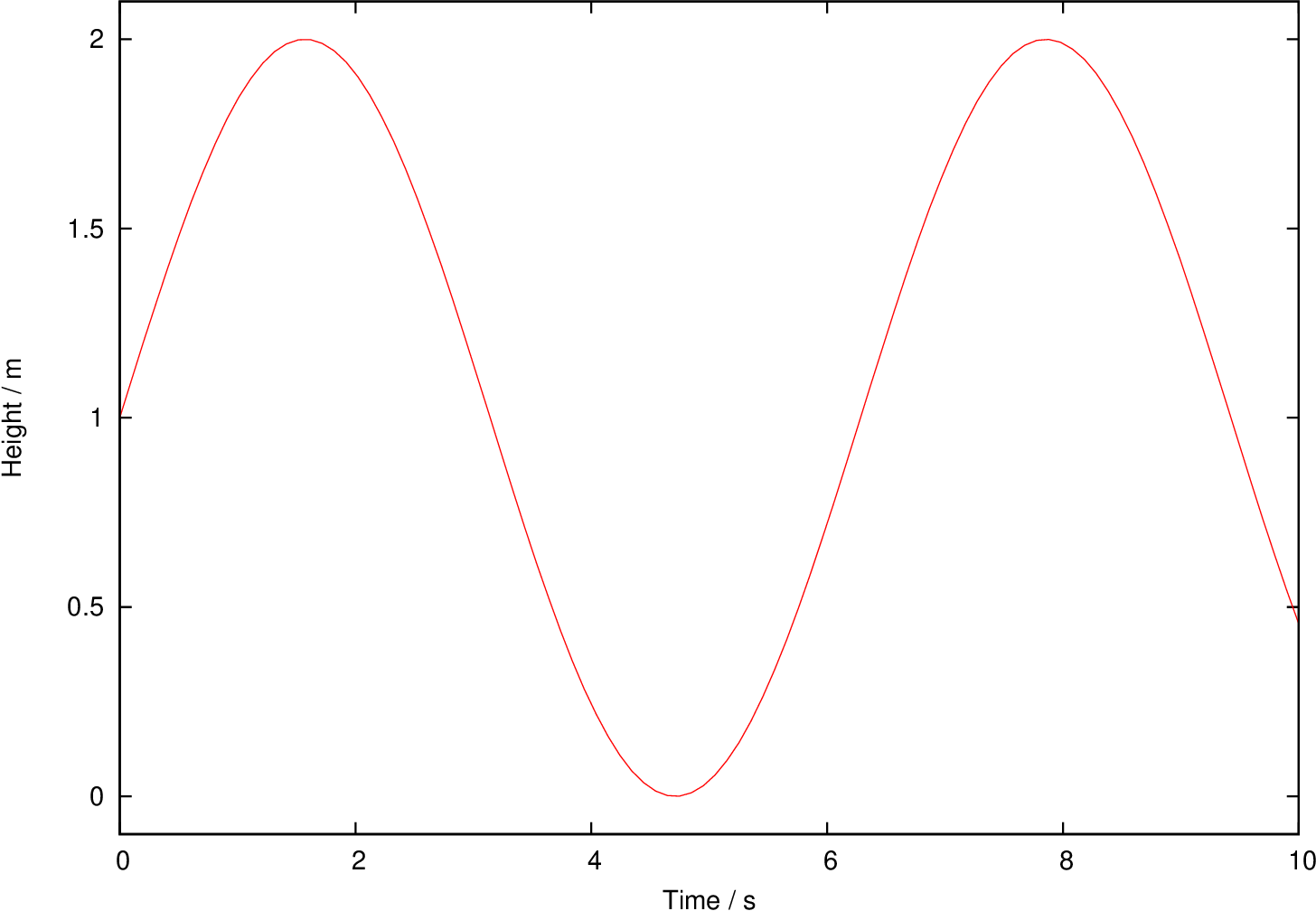}
    \caption{This is an example figure. Captions appear below each figure.
	Give enough detail for the reader to understand what they're looking at,
	but leave detailed discussion to the main body of the text.}
    \label{fig:example_figure}
\end{figure}

\begin{table}
	\centering
	\caption{This is an example table. Captions appear above each table.
	Remember to define the quantities, symbols and units used.}
	\label{tab:example_table}
	\begin{tabular}{lccr} 
		\hline
		A & B & C & D\\
		\hline
		1 & 2 & 3 & 4\\
		2 & 4 & 6 & 8\\
		3 & 5 & 7 & 9\\
		\hline
	\end{tabular}
\end{table}

\section{Conclusions}

The last numbered section should briefly summarise what has been done, and describe
the final conclusions which the authors draw from their work.
\end{instructions}

\section{Introduction}
\label{sec:Intro}
Pulsar timing, which has developed over more than five decades 
, is a sophisticated tool for tracking the rotational phase of a pulsar. Pulsars, owing to their highly stable rotation periodicities, are excellent clocks. The technique of pulsar timing has been successfully used in tests of General Relativity \citep[e.g.][]{1982ApJ...253..908T,2021PhRvX..11d1050K}, studying the properties of turbulent interstellar medium (ISM) \citep[e.g.][]{1990ApJ...349..245C}, and studying the interiors of some of the highest-density objects in the Universe \citep[e.g.][]{2010Natur.467.1081D,2013Sci...340..448A}. One of the most sought-after applications of pulsar timing is the detection of a stochastic gravitational wave background (GWB), of which the most probable source is thought to be the cosmic history of inspiralling supermassive black-hole binaries \citep{1990ApJ...361..300F,1995ApJ...446..543R}. With this as a primary goal, Pulsar Timing Arrays (PTAs) time a group of millisecond period pulsars (MSPs) spread across the sky. The presence of a GWB is expected to induce a common red noise process with spatial correlations among the pulsars \citep{1983ApJ...265L..39H}. 
The three major PTAs,  viz.  the European Pulsar Timing Array \citep[EPTA;][]{2016MNRAS.458.3341D}, the North American Nanohertz Observatory for Gravitational Waves \citep[NANOGrav;][]{2013CQGra..30v4008M} and the Parkes Pulsar Timing Array \citep[PPTA;][]{2008AIPC..983..584M} have been timing MSPs for almost two decades. These groups along with the Indian Pulsar Timing Array \citep[InPTA;][]{2022PASA...39...53T} are also working together in a joint collaboration, the International Pulsar Timing Array, to search for GWB in the combined data \citep[IPTA;][]{2022APS..APRG07003M}. The MeerKAT Pulsar Timing Array \citep[MPTA;][]{2023MNRAS.519.3976M} is an emerging PTA that is also contributing to the IPTA GWB searches.
Recently, multiple PTAs in their latest data release have reported a strong evidence for a statistically similar signal and marginal to strong evidence for a spatially correlated signature conforming to Hellings-Downs relation in their data set. A consistent characterstic strain amplitude of $A_{\rm yr} \sim2.4\times10^{-15}$ has been reported by PTAs  \citep{2023ApJ...951L...8A,2023arXiv230616214A,2023ApJ...951L...6R}, assuming a strain spectrum of the form $h_c(f) = A_{\rm yr} (f / 1\,{\rm yr}^{-1})^{2/3}$. For a confident detection of a GWB it is important that all the significant noise processes are accurately modelled.

It has been known since the discovery of pulsars that although the average pulse profile of most pulsars are stable, the individual pulses differ in shape and amplitude both with time and frequency. 
In frequency, some of the variation can be ascribed to external factors such as turbulence in the ISM, which causes increased pulse scatter broadening at lower frequencies \citep[e.g.][]{1974MNRAS.166..499W}. 
In time, much of the variation is intrinsic to the pulsar, and can contain clues to the radio emission mechanism in the pulsar magnetosphere. This phenomenon is commonly referred to  as "jitter". Studies of pulse jitter initially focused on slow pulsars ($P\sim 1$\,s), for which individual pulses are often higher in signal to noise (S/N) and hence easier to study \citep[e.g.][]{1985ApJS...59..343C}. However, it is now known that MSPs also exhibit jitter \citep{10.1111/j.1365-2966.2011.19578.x,2014MNRAS.443.1463S,2019ApJ...872..193L}. 

Crucially, the jitter in pulsars appears to manifest as a temporally uncorrelated random variation in pulsar timing signals. The random effect due to jitter is not accounted for in the standard algorithms for pulse time of arrival (TOA) measurement. Arrival times are usually estimated using the matched-filtering approach, which determines a best-fitting value of phase difference $\phi$ causing a time delay of $t_{\phi}$ by matching the averaged pulse profile $P(t)$ to a scaled version of a template $O(t)$ in the presence of a zero mean Gaussian radiometer noise $N(t)\sim\mathcal{N}(0,\sigma^2)$ with variance $\sigma^2$ and baseline offset $B$ \citep{1992PTRSL.341..117T} :
\begin{equation}
    P(t) = a O(t-t_{\phi}) + B + N(t)
    \label{eq:Profile model}
\end{equation}

The radiometer noise imposes a fundamental source of uncertainty in pulsar timing, which scales as
\begin{equation}
    \sigma_{\rm{toa}} \propto \frac{\rm{T_{Sys}}}{\rm{A_{eff}}}(\tau N_{\rm{p}}\Delta f)^{-\frac{1}{2}},
    \label{eq:radiometer noise}
\end{equation}
where $\rm{T_{sys}}$ is the system temperature of the receiver, $\rm{A_{eff}}$ is the effective area of the telescope, $\tau$ is the integration time, $N_{\rm{p}}$ is the number of polarisations and $\Delta f$ is the bandwidth over which the TOA was measured. Due to the phenomenon of jitter, the finite time averaged pulse profile $P(t)$ differs slightly from the template $O(t)$ \citep{1975ApJ...198..661H,1998ApJ...498..365J}. This difference varies stochastically at a single pulse timescale giving rise to an additional source of uncertainty in the TOA estimation process that is not accounted for in the formal arrival-time uncertainty from the fit. 
The excess uncertainty is referred to as "jitter noise" and it is usually modelled to  add in quadrature \citep{2015ApJ...813...65N} to the uncertainties expected due to the radiometer term. 

An effective way to increase sensitivity to gravitational waves is to time a larger number of pulsars each with higher precision \citep{2013CQGra..30v4015S}. For this reason PTAs continue to increase the size of their array as it is appropriate. 
The need of improving precision motivates undertaking PTA experiments on more sensitive radio telescopes.
The forthcoming Square Kilometer Array (SKA) is such a facility. Due to its increased collecting area and low $\rm{T_{sys}}$, the SKA is expected to have higher `timing potential' which is suggestive of as much as seven times improvement in the timing precision compared to other telescopes \citep{2013CQGra..30v4011L}. The SKA precursor telescope MeerKAT already benefits from having a larger collecting area and lower $\rm{T_{sys} \sim 18 K}$  \citep{2020PASA...37...28B} than {\em Murriyang} ($\rm T_{sys} \sim 22 K$), the  64-m Parkes radio telescope \cite[][]{2020PASA...37...12H}. Because of its location and observing efficiency, the MPTA is able to regularly observe 89 MSPs with declination $\delta < +30^\circ$   \citep{2023MNRAS.519.3976M}. With 4 years of data the MPTA is expected to contribute significantly to the international effort of GWB detection \citep{2022PASA...39...27S,2022APS..APRG07003M}. 

However, when pulsars are observed in the high signal to noise (SNR) regime, jitter noise dominates and limits the achievable precision \citep{10.1111/j.1365-2966.2011.19578.x,2014MNRAS.443.1463S,2019ApJ...872..193L}. This provides motivation for a better understanding of the jitter phenomenon and developing better models to account for it. \citet{2010arXiv1010.3785C} related the strength of jitter noise to the profile pulse width and introduced the jitter parameter $f_{J}$. Later using Parkes observations, \citet{2014MNRAS.443.1463S} confirmed that pulsars with jitter noise showed pulse-pulse shape variability and that the jitter noise in pulsar scales as $1/\sqrt{n}$ where $n$ is the total number of pulses integrated. Thus in the literature, the jitter noise is usually reported by assuming one hour of integration. Further, using the NANOGrav 12.5 year data set \citet{2019ApJ...872..193L} have found radio frequency dependence in the strength of jitter noise from many pulsars. Later \citet{10.1093/mnras/stab037} studied single pulse variability of jitter dominated millisecond pulsars and reported varying levels of jitter noise in their sample with values as high as 100 ns with one hour of integration.

While jitter noise is thought to be related to single pulse phase variations modulated by pulse intensities, there have been only a few joint studies of the two phenomena. 
This is due to the small sample of MSPs for which individual pulses can be detected and the high data rates necessary to record high time resolution observations necessary to temporally resolve the pulses. 

The brightest MSP and the subject of our work, PSR J0437$-$4715, has been studied in the most detail \citep{2012MNRAS.420..361L,2014MNRAS.441.3148O}. A few other MSPs have also had significant single-pulse studies. For example, 
PSR~J1713+0747 was observed to show amplitude dependence of pulse times of arrival \citep{Shannon_2012}. Later a similar effect was also reported  by \citet{10.1093/mnras/stab037} for PSR~J1909-3744. Further, \citet{10.1093/mnras/stab037} also detected evidence of fluence variability in the pulses of J1909$-$3744, which they ascribed to pulse nulling. In a following study of the single pulses emitted by J1909$-$3744, \citet{2022MNRAS.510.5908M} determined this was in fact a mode-changing phenomenon and showed improvement in the timing precision when only considering the more energetic mode. Study of single pulses has also led to new techniques of improving timing precision. Recently, \citet{2023arXiv230402793N} have explored the use of a dynamic pulse fitting technique to improve the precision of timing in a mode changing pulsar.

While techniques have been developed to study and mitigate temporal variability attributed to pulse jitter, there have been fewer investigations into the spectral impact of pulse-shape variations. 
PSR J0437$-$4715 is the nearest and brightest MSP, which is timed regularly by PTAs, and is highly jitter dominated. This makes it the best candidate for such a study. For this pulsar, \citet{10.1093/mnras/stab037} have measured the jitter noise to be 50 ns with one hour of integration. Although \citet{10.1111/j.1365-2966.2011.19578.x} showed a high correlation between TOAs calculated at frequencies within a bandwidth of 250MHz, later wide-band studies have identified decorrelation in jitter when TOAs were separated by a larger fractional bandwidth. This was shown to occur in observations separated by 2 GHz using the Parkes radio telescope \citep{2014MNRAS.443.1463S}. \citet{10.1093/mnras/stab037} additionally reported that the variance of the jitter noise in J0437$-$4715 has a frequency dependence with higher variance at lower frequencies.  In addition, jitter in MeerKAT L- band observations was seen to be decorrelating  across a bandwidth of 856MHz \citep{10.1093/mnras/stab037}. A similar effect has been identified in PSR J1713$+$0747 \citep{2016ApJ...819..155L}. 

While searching for a spatially correlated GWB signal, PTA experiments are required to accurately account for additional stochastic signals. 
Jitter noise is no exception; it is included as a white (temporally uncorrelated) noise parameter, termed ECORR in most pulsar-timing software. This parameter was first introduced by \citet{2015ApJ...813...65N} to incorporate temporally uncorrelated but  spectrally correlated errors.
Importantly,  the ECORR model assumes that the noise is completely correlated across radio frequencies and does not have any frequency dependence. Previous work on PSR J0437$-$4715 has shown that both these assumptions are not valid. Apart from that, the ECORR parameter has also been observed to have a larger amplitude than predicted from  jitter noise \citep{2016ApJ...819..155L} suggesting  inaccuracy or misspecification in its modelling. Although the jitter decorrelation is likely to be of less importance when the data are taken in a relatively narrow bandwidth, PTAs are shifting towards use of ultra wideband receivers. For example, the current data release of the PPTA \citep[][]{2023PASA...zic} has used the newly installed Ultra Wideband Low \citep[UWL;][]{2020PASA...37...12H} receiver, which has a wide frequency coverage from $\sim$700 MHz to 4000 MHz. 
As  next-generation telescopes such as FAST and the SKA become operational, more pulsars are expected to be sufficiently bright for the single pulse studies and for their jitter noise characterisation \citep{2011MNRAS.417.2916L,10.1111/j.1365-2966.2011.19578.x}. From this point of view, a more complete white-noise model with an accurate description of jitter noise is needed, as model inaccuracies would leave an impact on PTA accuracy and precision.

Motivated by this challenge, we have studied in detail the decorrelation of jitter noise in J0437$-$4715 and have constructed a consolidated model which accounts for the decorrelation and frequency dependence. We show that the unmodelled frequency dependence in white noise biases other chromatic processes such as long term dispersion measure (DM) variations. Section \ref{sec:Data}, \ref{sec:Method} together describe the data reduction process for our analysis. Section \ref{sec:Result&Disc} describes results in detail. Section \ref{sec:updating_jitter} describes our method of incorporating the decorrelation in the PTA noise analysis.


\section{Data Reduction}
\label{sec:Data}
The decorrelation in jitter noise becomes increasingly clear by analysing TOAs measured with wider frequency separations and greater sensitivity.  MeerKAT is one of the best existing telescopes for such studies. The array is located in the Great Karoo region of South Africa and the telescope site is relatively free of radio frequency interference (RFI). The L-band receiver of MeerKAT has a low system temperature $\rm{T_{Sys}}$ of $\sim$ 18K, a gain of 3  K $\rm Jy^{-1}$ with a usable bandwidth of $\approx$ 780 MHz centred at 1283\,MHz \citep{Bailes_2020_PASA}. Thus to perform the wideband study of jitter we worked primarily with the data from MeerKAT L-band receiver.

Our MeerKAT data were 4 minutes in duration with a frequency spanning from 856 to 1712 MHz, and was recorded on MJD 59718 in raw voltage format as a part of MeerTime program \citep{2016mks..confE..11B}. The data were first polarization and flux calibrated in the correlator and then written to disk.  The \textsc{DSPSR} \citep{2011PASA...28....1V} software package was used to coherently dedisperse and detect $\sim$ 44,000 single pulses from this observation. These single pulse data products contained 1024 frequency channels across the entire bandwidth and 1024 bins of phase resolution. To avoid the filter roll-off of the receiver, 96 channels were removed (48 from each band edge), reducing the number of channels to 928. A median smoothed difference algorithm of the tool \texttt{paz} from the software package \textsc{psrchive} \citep{2004PASA...21..302H} was used to excise RFI. Finally the data were frequency averaged into 32 frequency channels.

We complemented our MeerKAT observations with a 4 minutes observation recorded with Murriyang on MJD 59791 using the UWL receiving system, which has a $3300$\,MHz bandwidth spanning $732$ to $4032$\,MHz. 
 These data were recorded in pulsar search mode and were further processed using \textsc{DSPSR} to dedisperse and generate single pulse data products. After cleaning RFI, the data were flux and polarization calibrated by the \textsc{psrchive} program \texttt{pac} using a fluxcal file and  a noise diode observation which was recorded along with the pulsar data. Single pulses were then saved with 64 frequency channels and 1024 phase bins. 

Jitter is a single pulse phenomenon. Recording single pulses allows us to study the evolution of jitter noise and its decorrelation as we integrate a higher number of pulses. To understand this effect we generated data sets containing progressive number of pulses integrated, starting from 4 pulses per integration, up to 1024 pulses per integration with steps of powers of two. This pulsar shows a considerable amount of frequency evolution in its pulse profile, as it broadens towards lower frequencies. Since our data from both telescopes have a wide bandwidth, using a frequency averaged template for generating TOAs may introduce systematic biases in the TOA estimation. Thus to keep our analysis free from this confusion, we used a frequency dependent 2D template (also called a portrait) of this pulsar produced using the software \textsc{PulsePortraiture} \citep{2019ApJ...871...34P}. Finally the sub-banded TOAs were calculated for each data set using the \textsc{psrchive} command \texttt{pat}.
Before finalising the TOA data sets, we used \textsc{tempo2}'s \texttt{plk} plugin \citep{2006MNRAS.369..655H} to visually assess the quality of data. TOAs from some frequency channels were found to be affected by RFI and were manually deleted from the data set. 

Finally to test the main outcome of our analysis, we also used a MeerKAT four-year pulsar timing data set collected as the part of MPTA program. These data were recorded at the L-band using the PTUSE backend \citep{Bailes_2020_PASA} and comprised integrations of $\sim 256$ seconds at each epoch. The data were coherently de-dispersed and calibrated using the custom pipeline \textsc{meerpipe}. The RFI excision was performed using the program \textsc{meerguard} which is a modified version of \textsc{coastguard} \citep{2016MNRAS.458..868L}. These data were frequency averaged to 16 channels across the L-band and the TOAs were generated using the \textsc{psrchive} utility \texttt{pat}. These data were used to compare the performance of the new jitter noise model developed during the present work against the existing jitter noise model.

\section{Methodology}
\label{sec:Method}
We started our analysis with the 4 minutes duration high time resolution data sets comprising observations of increasing integrations from both the telescopes. The first step was to produce residuals by subtracting the best known timing model from the observed TOAs. A long-term timing ephemeris was used which was based on that from the first data release of MPTA \citep{2023MNRAS.519.3976M}. We used \textsc{tempo2} to obtain residuals by fitting for DM and pulse period to the entire 4 minutes data set. For this fit all the binary parameters were held fixed as more precise measurements were obtained previously by PPTA (D.Reardon et al. in preparation). The resultant residuals were saved to disk for further analysis. The covariance matrix of these output residuals contains contribution from the radiometer term as well as jitter term. The jitter contribution can be identified using relation 
\begin{equation}
    {\rm{\sigma_{TOA}}}^2 = {\rm{\sigma_{S/N}}}^2 + {\rm{\sigma_{j}}}^2 ,
    \label{eq:Sigma_jitter}
\end{equation}
where ${\rm{\sigma_{TOA}}}^2$ represents the observed covariance in the residuals, ${\rm{\sigma_{S/N}}}^2$ represents the contribution because of the radiometer noise in the receiver, ${\rm{\sigma_{j}}}^2$ represents the contribution due to jitter noise. The covariance matrix is a crucial diagnostic in analysing any stochastic signal. We based the majority of our analysis on the jitter covariance matrix of  data sets of increasing integration. In order to estimate the ${\rm{\sigma_{S/N}}}^2$, we performed simulations using plugins available in \textsc{tempo2}. First we generated ideal TOAs based on our best timing model by using the \texttt{formIdeal} plugin. Then a Gaussian noise was added to the ideal TOAs, such that the simulated uncertainties matched those due to radiometer noise. This was done using the \texttt{addGaussian,} and \texttt{createRealisation} plugins. The covariance matrix of these ideal TOAs is diagonal and must be subtracted from the covariance matrix of the residuals to obtain the jitter covariance matrix. This method was previously shown to accurately measure the level of jitter noise by \citet{2021MNRAS.502..407P}
We performed this operation on all our data sets and then the resultant matrices were used for further analysis which is described in detail in section \ref{subsec:Jitter_Covariance_matrix}.

The stochastic nature of the residual profiles, which are the difference between our template and the true pulse profile, gives rise to jitter noise in pulsars. This implies that any characteristic behaviour of jitter noise inferred from TOAs should be  directly traceable to the residual profiles themselves. As such, we also analysed the profile residuals to relate the decorrelation to the pulse energy variations.
We performed this analysis by using the energy statistic computed in the units of signal to noise ratio which is given as
$$S/N= \frac{\sum_{\rm i=1}^{\rm{N_{window}}} (A_{\rm i} -B)}{\sqrt{\rm{N_{window}}}\rm{\sigma_{off}}}$$ on the profile residuals. 
Here $A_{\rm i}$ is the intensity at $\rm i^{th}$ bin in the residual profile, $B$ is average intensity of off pulse region, $\rm{\sigma_{off}}$ is rms in the off-pulse region. $\rm{N_{window}}$ is the number of bins summed together. 

Before computing the energies we first de-dispersed the entire data set with the DM obtained from a fit performed using \textsc{tempo2} in order to remove excess  dispersion induced contributions to the residual profiles. We then obtained 
 the phase-aligned and scaled frequency-dependent version of the template by fitting the original template to the frequency dependent profile averaged over entire observation. Then the scaled template was subtracted from the single pulse profiles to obtain residual profiles. 
Using these residual profiles, a two-dimensional array containing energies in the residual profiles as a function of frequency was computed.
The energies thus calculated had zero mean value and they contained contributions from pulse shape variations and the inherent noise in the telescope. To subtract the latter, energies were computed in the off-pulse region using the same relation. 
Finally, the covariance matrices were constructed as described earlier from these energies and were used for further analysis. 

In both of the above mentioned analyses we do not consider the effect of interstellar scintillation from the ISM as the scintillation time scale for PSR J0437-4715 is $\sim$ 40 minutes \citep{2020ApJ...904..104R}.  Our observations span a much shorter timescale of 4 minutes. Additionally our sub-banded data have a bandwidth of $\sim$ 25 MHz; while the scintillation bandwidth $\Delta \nu_{d}$ at the L band centre frequency was calculated to be $\sim$ 160 MHz from the reference value reported by \citet{2020ApJ...904..104R}

\section{Results}
\label{sec:Result&Disc}
We present our analysis in three parts. We first analyse the jitter covariance matrices. We then obtain an empirical relation to characterise the rate of decorrelation. Finally we extend our study by performing the spectral analysis on the jitter covariance matrix and use it to improve the white noise modelling.
\subsection{Analysis of jitter covariance matrix}
\label{subsec:Jitter_Covariance_matrix}
\begin{figure}
	\includegraphics[width=\columnwidth]{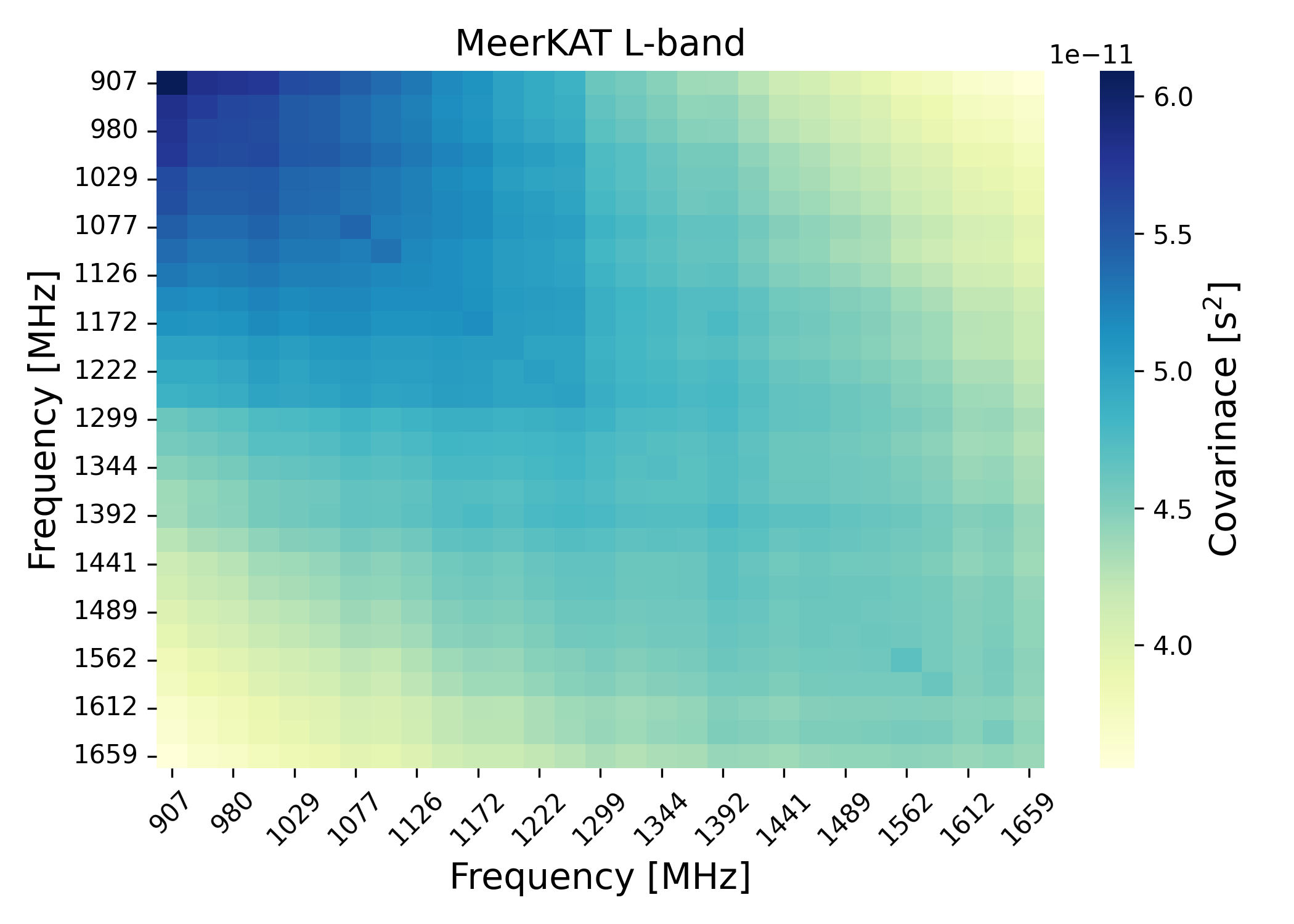}
    \caption{Jitter covariance matrix obtained after subtracting the radiometer noise contribution from 32 pulses integrated L-band data from MeerKAT.  }
    \label{fig:Covariance_mat_J0437}
\end{figure}

\begin{figure*}
    \begin{subfigure}[t]{.49\linewidth}
        \centering\includegraphics[width=1\linewidth]{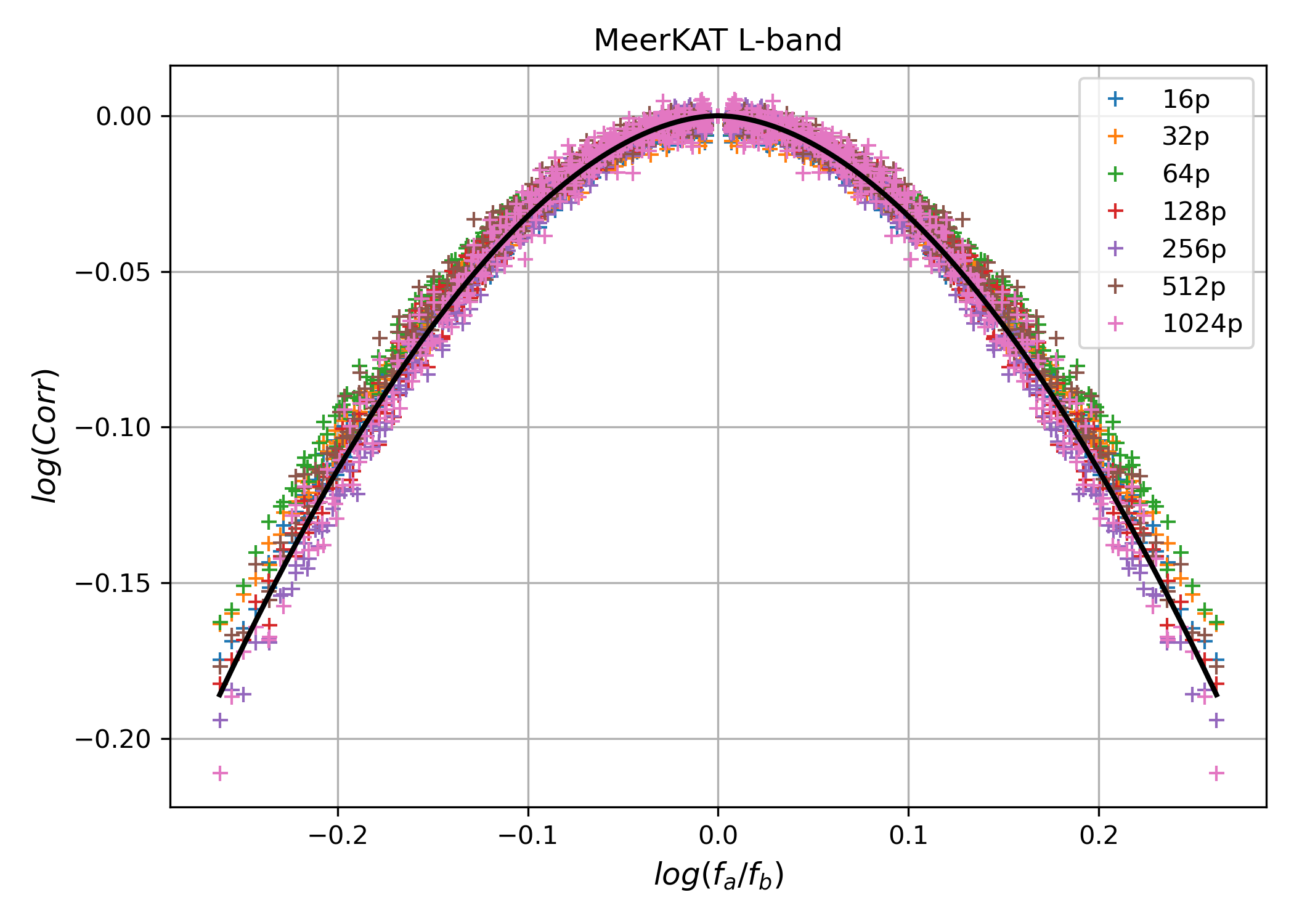}    \caption{}
    \label{fig:Correlation_arcs_J0437} 
    \end{subfigure}
    \begin{subfigure}[t]{.49\linewidth}
        \centering\includegraphics[width=1\linewidth]{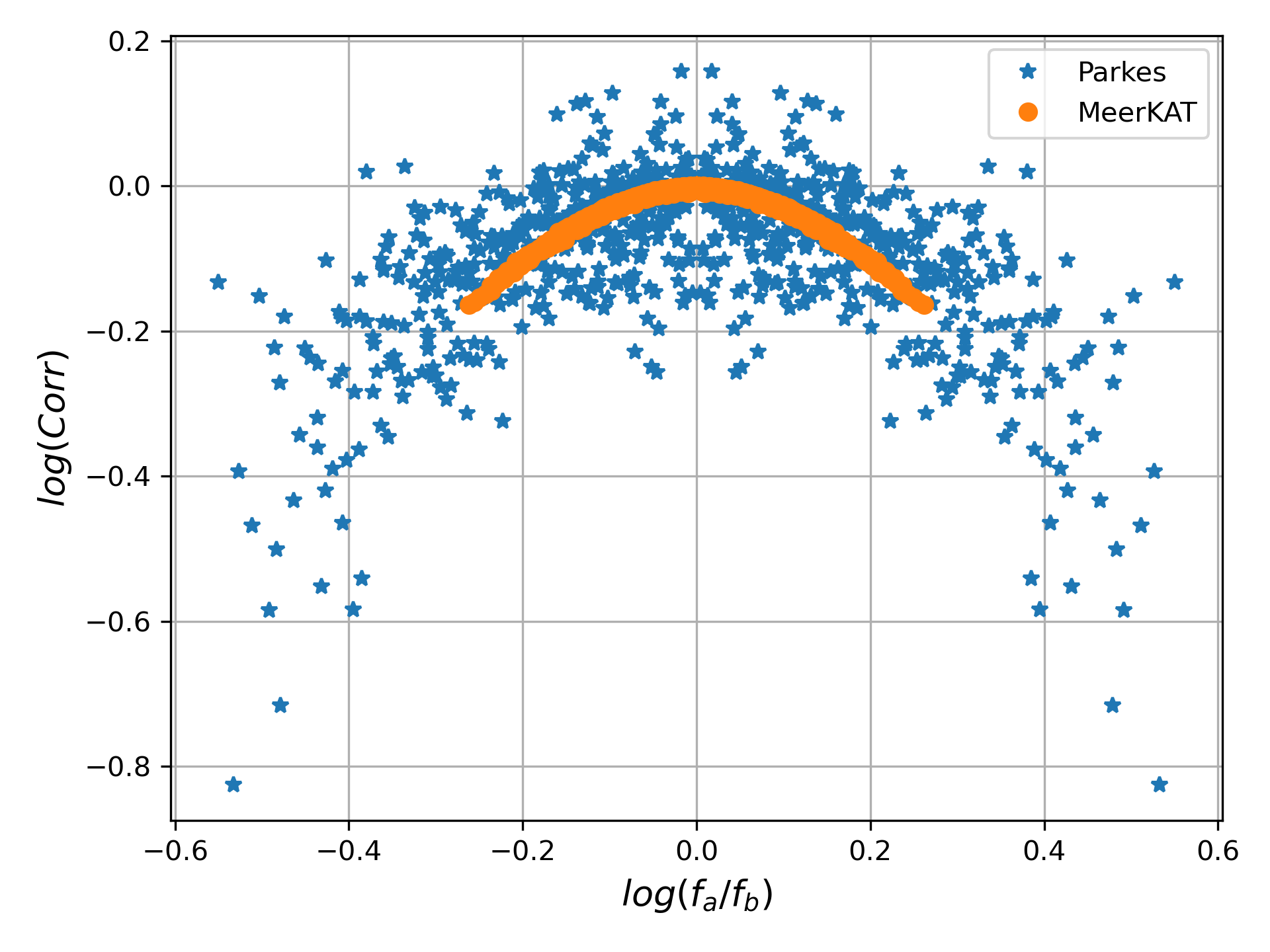}
            \caption{}
    \label{fig:pks_mk_CorrArcs_compare}
    \end{subfigure}\\
                \caption{\textbf{Panel a}: Jitter correlation coefficients plotted for the data containing increasing amount of single pulses integrated. Self similarity in the decorrelation mechanism can be inferred as the arc does not depend explicitly on the values of frequencies $f_{a}$ and $f_{b}$. The black solid line shows the Power-law fit with parameter values $\alpha = 1.82 \pm 0.01$ and $\beta = -2.13 \pm 0.04$.
            \textbf{Panel b}: Comparison of correlation arcs obtained from MeerKAT and Parkes data: The Parkes data show a wide scatter due to its low signal to noise ratio per unit bandwidth. }
\end{figure*}

When considering timing covariances,  we identified that integrating to a higher number of pulses preserves the overall form of the covariance matrix. However, the absolute values of covariance scale as $\sqrt{n}$ for all frequencies, as expected for a white-noise process\
.  To illustrate this, a covariance matrix with the 32-pulse integrated data set is shown in Fig.~\ref{fig:Covariance_mat_J0437}. It can be seen from the matrix that the diagonal elements (variance at each sub-band) drops monotonically. This shows that the jitter noise has a frequency dependence. This was also reported by \citet{2021MNRAS.502..407P} and \citet{2014MNRAS.443.1463S} for this pulsar and by \citet{2019ApJ...872..193L} for pulsars in NANOGrav 12.5 year data set. Secondly the decrease in covariance values as one departs from the diagonal along a row/column has a similar trend regardless of the frequency. This trend becomes clear when the correlation coefficients at any two frequencies are computed using equation~\ref{eq:Correlation_Coefficient}. 
\begin{equation}
    \rho(f_{a},f_{b})=\frac{cov(f_{a},f_{b})}{\sqrt{cov(f_{a},f_{a})cov(f_{b},f_{b})}}
    \label{eq:Correlation_Coefficient}
\end{equation} The previously observed de-correlation in the jitter noise was once again confirmed in this analysis. Moreover a pattern was seen when the logarithm of correlation coefficient is plotted against the logarithm of ratio of two frequencies $(f_{a},f_{b})$ as shown in Fig.~\ref{fig:Correlation_arcs_J0437}. It can be seen that regardless of values of  $f_{a}$ and $f_{b}$, the correlation coefficient can be described by an arc. This demonstrates the inherent self similarity of the de-correlation. The arc can be modelled as a power-law function:
\begin{equation}
    log(\rho(f_{a},f_{b})) =  \beta 
    \left|\left|log\left(\frac{f_{a}}{f_{b}}\right)\right|\right|
    ^\alpha ,
    \label{eq:Jitter_freq_dependence}
\end{equation}
where $\alpha$ and $\beta$ are parameters with best-fitting values as $\alpha = 1.82 \pm 0.01$ and $\beta = -2.13 \pm 0.04$. Fig~\ref{fig:Correlation_arcs_J0437} also confirms that averaging pulses does not have any significant effect on structure or curvature of the arc. The observed arc suggests that TOAs separated by span of 775MHz at MeerKAT L-band have a correlation coefficient of just 0.67. Fig 3 from \citet{2021MNRAS.502..407P} shows the Spearman correlation coefficient calculated on 8-sec integrated data and the value can be seen to be lower than what we report. We attribute the difference to the type of correlation coefficient reported. 
We find a consistent value of 0.5 $\pm$0.2 when measuring the Spearman correlation coefficient on our 1024 pulse integrated data for frequencies maximally separated in our data set.
The presence of a correlation arc is also confirmed in the data collected using Parkes UWL receiver. Fig~\ref{fig:pks_mk_CorrArcs_compare} shows the comparison between the data obtained from the two telescopes.

\begin{figure}
	\includegraphics[width=\columnwidth]{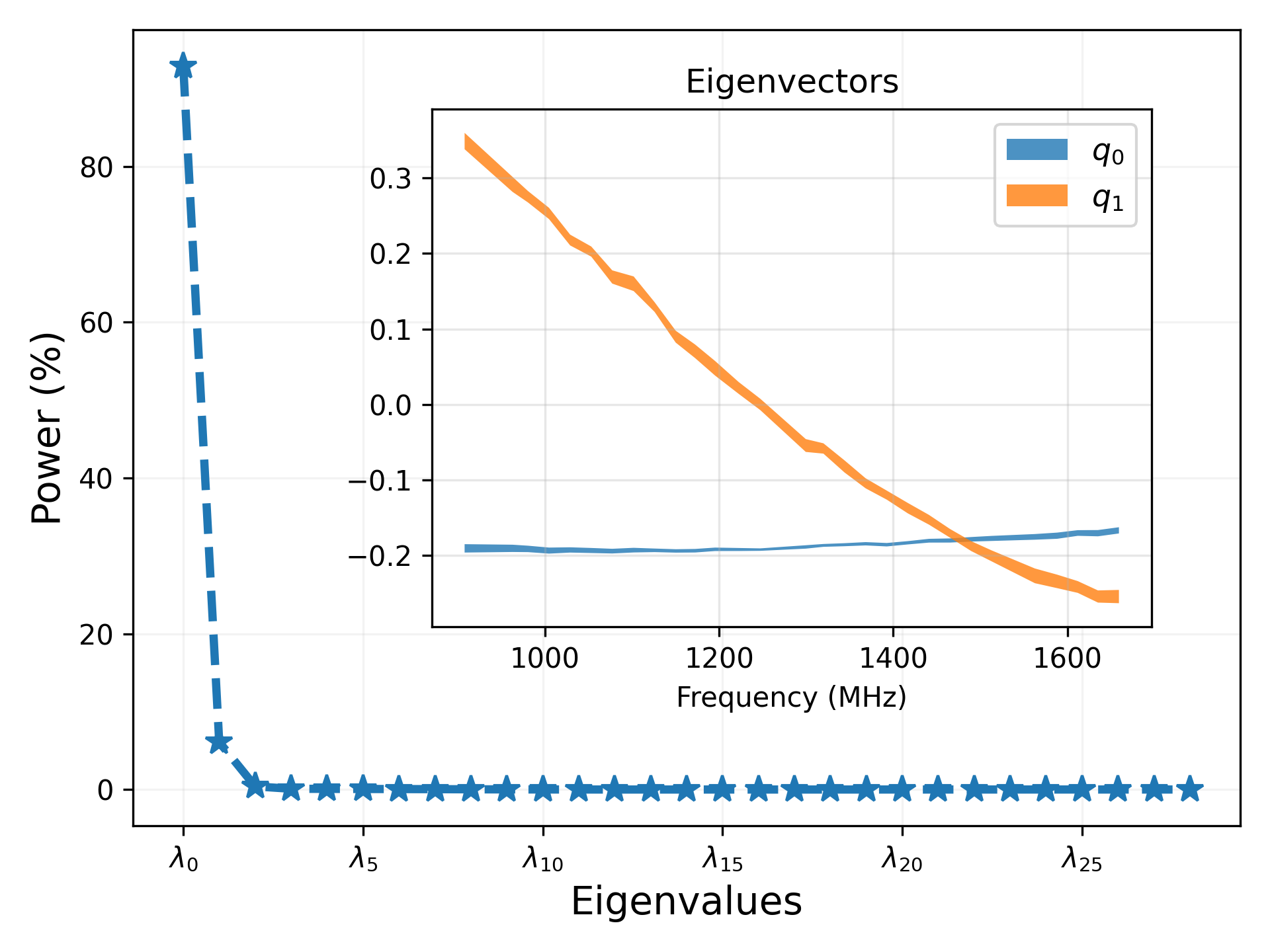}
    \caption{Output of singular value decomposition of jitter covariance matrix. \textbf{Main panel}: shows the relative strengths of eigenvalues in ascending order. \textbf{Inset panel}: shows the frequency dependence of eigenvectors $\vec{q_{0}}$ and $\vec{q_{1}}$. We performed bootstrapping to obtain the uncertainties on the eigenvectors. The shaded region represents the 95\% confidence region.}
    \label{fig:SVD_covariance_mat_J0437}
\end{figure}

A further mathematically rigorous analysis was performed by finding the singular value decomposition (SVD) of the covariance matrix obtained using data from MeerKAT L-band receiver. SVD is a powerful method which decomposes a matrix into several rank-1 matrices. In the case of symmetric matrix $\text{J}$, such as a covariance matrix, the SVD can be expressed as 
\begin{equation}
\begin{split}
    J &= Q \Lambda Q^T = 
    \begin{bmatrix}
    & & \\
    \vec{q_{0}} &\dots &\vec{q_{n}}\\
    & & 
\end{bmatrix}
\begin{bmatrix}
    \lambda_{0} & &\\
    & \ddots & \\
    & & \lambda_{n}
\end{bmatrix}
    \begin{bmatrix}
   & \vec{q_{0}}^T& \\
    & \vdots& \\
    &  \vec{q_{n}}^T&
\end{bmatrix}
\\
& = \lambda_{0}\vec{q_{0}}\vec{{q_{0}}}^T + \dots + \lambda_{n}\vec{q_{n}}\vec{{q_{n}}}^T
\end{split}
\end{equation}
Where $\lambda_{0}\ge\lambda_{1}\dots\ge\lambda_{n}$ are eigenvalues of the matrix $J$ in increasing order of their magnitude and $\vec{q_{0}} \dots \vec{q_{n}}$ are corresponding orthonormal eigenvectors. The relative magnitudes of eigenvalues indicate the importance of the corresponding eigenvectors in constructing the original matrix $J$. Fig.~\ref{fig:SVD_covariance_mat_J0437} shows the fractional contribution of the eigenvalues to the jitter covariance matrix. All the eigenvalues above the second  ($\lambda_{1}$) contribute negligibly, so we do not consider them.
Thus the jitter covariance matrix can be approximated to be $J \approx \lambda_{0}\vec{q_{0}}\vec{{q_{0}}}^T + \lambda_{1}\vec{q_{1}}\vec{{q_{1}}}^T$.  

The SVD analysis was not performed on the covariance matrix obtained from the Parkes UWL. Due to the lower sensitivity of the UWL relative to  the MeerKAT L-band, the data at higher frequencies have greater contribution from radiometer noise compared to jitter noise. The Parkes radio telescope site is more affected by RFI (especially across the broad band of the UWL receiver). The RFI can be missed by the filtering tools and can result in detection of higher significant eigenvalues.

Historically, the jitter noise in pulsars is assumed to be fully correlated across the band. As such  its covariance matrix can be modelled as a rank-1 matrix  $\rm J = ECORR^{2}\vec{u}\vec{{u}}^T$. In this case, $\rm ECORR^{2}$ is the only non-zero eigenvalue which parameterises the jitter noise and $\vec{u} = (1\dots1)^{T}$ is the only eigenvector. However PSR J0437$-$4715 exhibits jitter covariance matrix of rank-2 as there are at least 2 significant eigenvalues. The matrices produced by individual eigenvectors $\vec{q_{i}}\vec{q_{i}}^T$ are rank-1 each and hence show full correlation just on their own, however the de-correlation seen in Fig~\ref{fig:Correlation_arcs_J0437} arises due to addition of contributions from each of these components. The use of $\vec{u} = (1,\dots,1)^T$ as the only eigenvector in the original rank-1 definition implies no frequency dependence of jitter noise in pulsars. However as shown in Fig~\ref{fig:SVD_covariance_mat_J0437}, the eigenvectors of J0437$-$4715 do have a modest frequency dependence in $\vec{q_{0}}$ and a stronger frequency dependence in $\vec{q_{1}}$. The confidence intervals on the eigenvectors of \ref{fig:SVD_covariance_mat_J0437} were estimated by the method of bootstrapping from the 32 pulses integrated data set \citep{10.1214/aos/1176344552}.

\subsection{Decorrelation in pulse energy variations}
\label{subsec:Pulse_shape_Variations}
\begin{figure*}
    \begin{subfigure}[t]{.49\linewidth}
        \centering\includegraphics[width=1\linewidth]{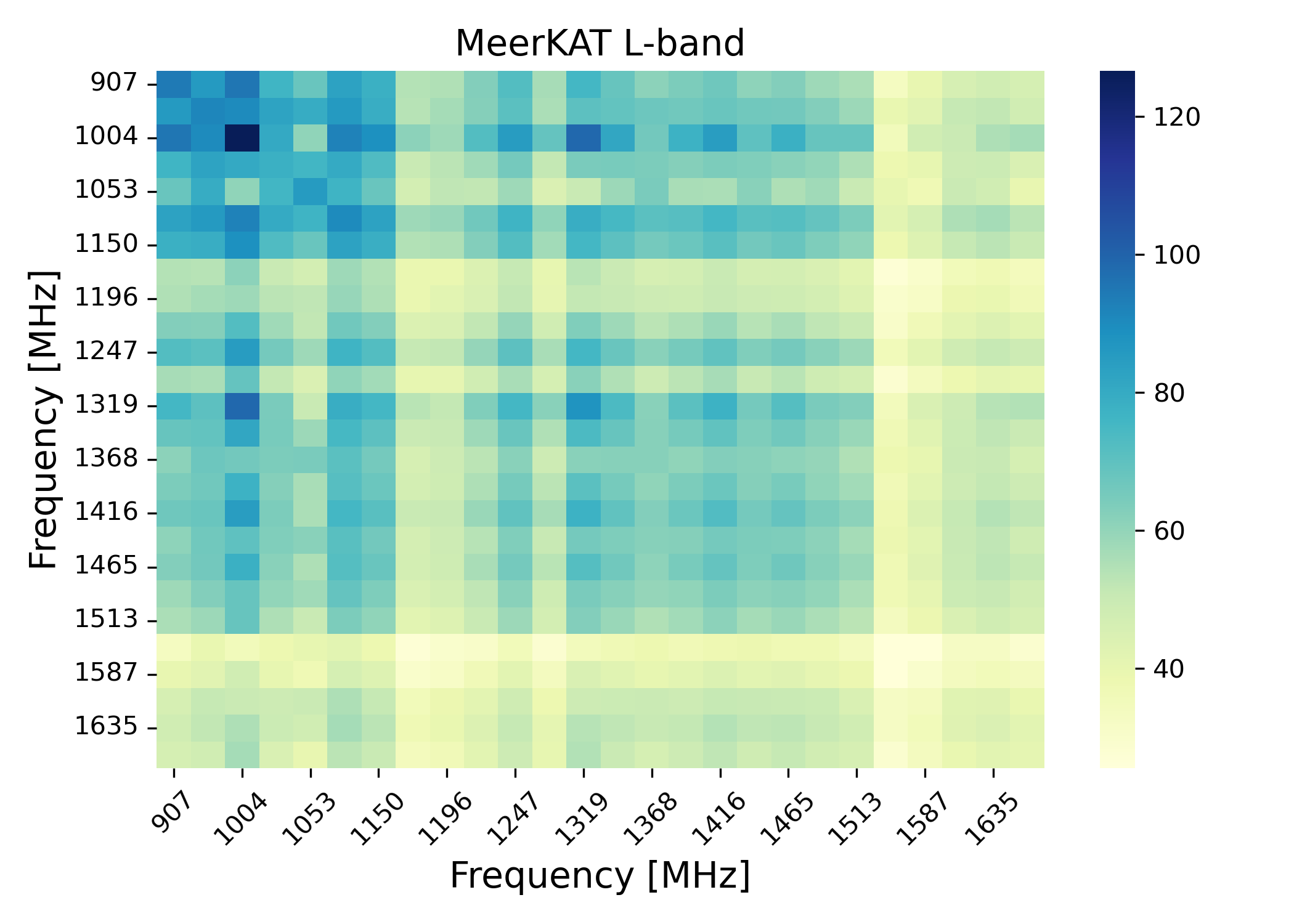} 
        \caption{}
        \label{fig:Phase_Covariance_mat_J0437}
    \end{subfigure}
    \begin{subfigure}[t]{.49\linewidth}
        \centering\includegraphics[width=1\linewidth]{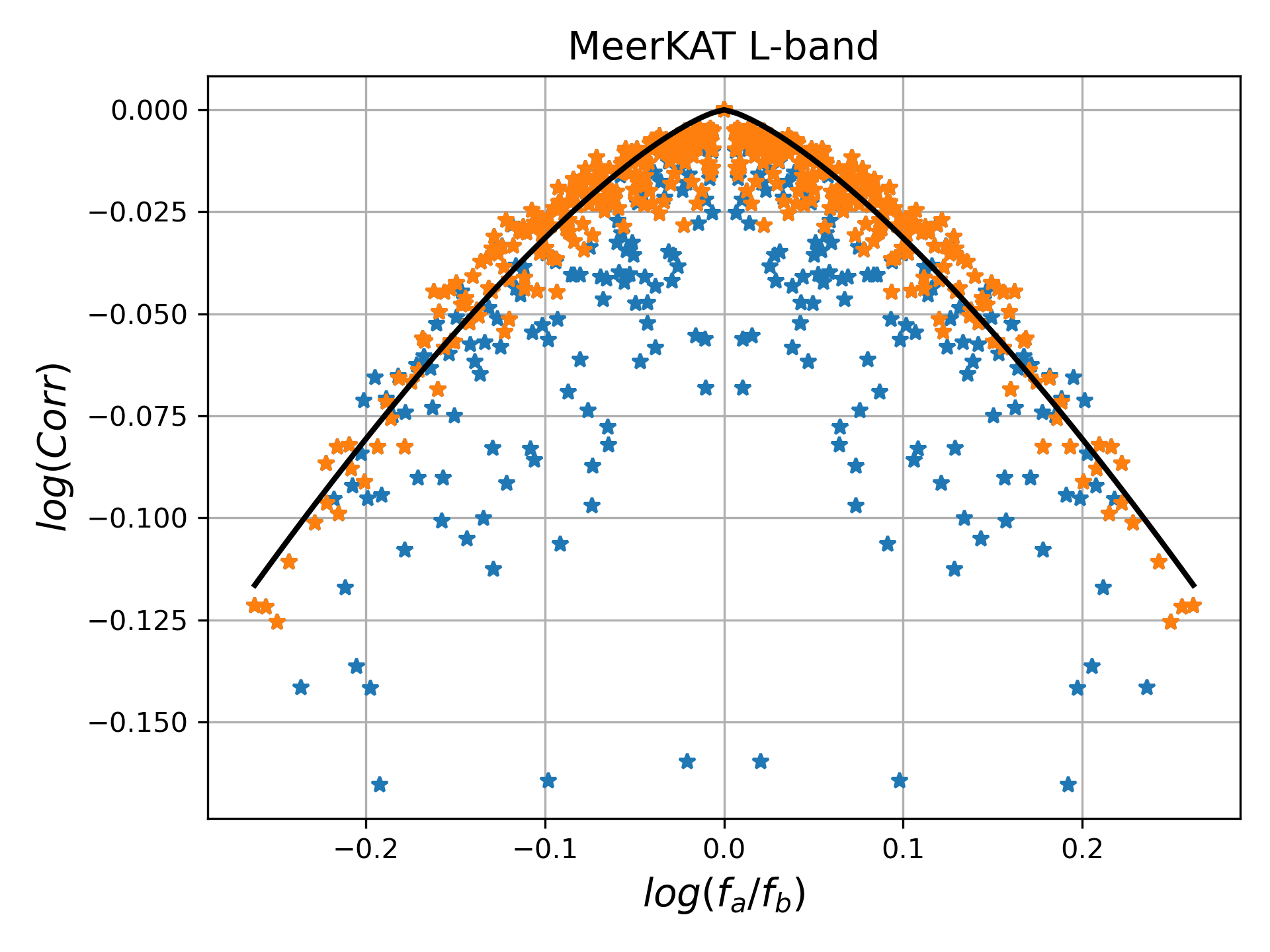}
        \caption{
        }
        \label{fig:Phase_Correlation_arcs_J0437}
    \end{subfigure}\\
    \caption{\textbf{Panel a}: Energy covariance matrix obtained from 32 pulse integration data. \textbf{Panel b}: Correlation arc obtained from the energies in residual profiles. The blue stars represent the correlation points at all frequency pairs. The orange stars are correlations obtained after excluding energies at some frequencies. The black solid line shows the power-law fit to the orange stars with parameter values $\alpha = 1.36 \pm 0.03$ and $\beta = -0.72^{+0.04}_{-0.03}$.
    }
    \label{fig:Phase_Cov_combined}
\end{figure*}

The aim of this part of our work is to relate the observed decorrelation in TOAs to the profile residuals as a function of frequency. We do this by computing the covariance matrix of the energies in the single pulse profile residuals according to the procedure described in section \ref{sec:Method}. As this covariance matrix captures the energy variations of single pulses, we refer to it as the energy covariance matrix. Similar to jitter noise covariance matrix, decorrelation was also seen in this observable. Fig~\ref{fig:Phase_Cov_combined} shows the covariance matrix and the correlation arc plotted according to equation~\ref{eq:Correlation_Coefficient}. It was noticed that the pulse energy variations at certain frequency channels showed significantly lower correlation than the trend followed by the data from rest of the frequency channels. It is suspected that this was caused due to weak RFI at those frequencies. Thus after excluding data from such frequency channels, a power-law of equation~\ref{eq:Jitter_freq_dependence} was fitted to the rest of data. The best fitting values obtained were $\alpha = 1.36 \pm 0.03$ and $\beta = -0.72^{+0.04}_{-0.03}$. It can be noticed that the degree of curvature of the pulse energy decorrelation arc is less than that of jitter decorrelation arc demonstrating that the pulse energies decorrelate to a lesser extent than the arrival times calculated from the pulses. Following from the jitter model given by \citet{2010arXiv1010.3785C}, the jitter noise is a result of variations in pulse shape modulated by variations in intensity. Thus, it is reasonable that the pulse energy decorrelation can only partially explain the jitter decorrelation. The remainder of the decorrelation may be ascribed to the pulse width variations.


\section{Implementing a Rank-2 model for jitter noise}
\label{sec:updating_jitter}
In any parameter estimation problem, the precision and accuracy of estimated model parameters requires a robust description of the signals and noise in the data. From this perspective, the decorrelation in jitter noise needs to be appropriately accounted for in pulsar timing analysis, in particular for GWB searches. 
The pre-fit timing residuals $\vec{\delta t}$ of a pulsar contain contributions from deterministic and stochastic signals. Variations in the timing model parameters give rise to deterministic signals, whereas the randomly varying signals like DM variations, spin noise, stochastic gravitational waves and jitter noise constitute the stochastic component. The post-fit residuals $\vec{r}$ can then be written as 
\begin{equation}
    \vec{r} = \vec{\delta t} - \mathbf{T}\vec{b} ,
    \label{eq:timing_residuals}
\end{equation}
where $\mathbf{T}$ is the design matrix and $\vec{b}$ is the vector of model parameters. The jitter noise can be handled in two different methods. 
In the first method jitter is included in the overall white noise covariance matrix $\mathbf{N}=\langle \vec{r} \vec{r}^T \rangle$ of post-fit residuals. The matrix $\mathbf{N}$ becomes a block-diagonal matrix and it is required to be inverted for calculating the likelihood function. For a rank-1 jitter covariance matrix the inverse can be computed using Shermon-Morrison formula \citep{2021arXiv210513270T}. 

The second method models jitter as a zero-mean Gaussian process which is uncorrelated across different epochs but is correlated within a given epoch. Under this framework jitter noise signal can be written as  
\begin{equation}
   \mathrm{ \vec{n}_{jitter} = \mathbf{U}\vec{j} },
    \label{eq:jitter_GP}
\end{equation}
where $\mathbf{U}$ is jitter basis matrix of dimensions $(N_{\rm{toa}} \times N_{\rm{epoch}})$  constructed out of eigenvector $\vec{u}$. Here, $\vec{j}$ is a jitter vector of length $N_{\rm{epoch}}$ which is parameterised by a single parameter $\rm ECORR$. The jitter basis matrix $\mathbf{U}$ is added to the design matrix $\mathbf{T}$ and the jitter vector $\vec{j}$ becomes a part of model parameters vector $\vec{b}$. Further description of this methodology can be found in \citet{2021arXiv210513270T}.

The software package \textsc{enterprise}
\citep{enterprise} is a Bayesian parameter estimation code that is commonly used by PTAs for noise modelling and searches for nanohertz-frequency gravitational waves. It uses the rank-1 definition of jitter noise for all pulsars.
It is clear from the analysis in sec \ref{subsec:Jitter_Covariance_matrix} that for PSR J0437$-$4715 both the eigenvalues along with their corresponding eigenvectors should be included in the white noise model and sampled during searches to improve the results. 

In order to understand the importance of using a rank-2 model, we modified the current rank-1 jitter implementation in \textsc{enterprise}. When jitter is included as part of the white noise covariance matrix, the key modification involves using the generalised Sherman-Morrison-Woodbury formula for a rank-k perturbation for matrix inversions 

\begin{equation}
    (A + UV^T)^{-1} = A^{-1} - A^{-1}U(I + V^{T}A^{-1}U)^{-1}V^{T}A^{-1} ,
    \label{eq:ShMorWo}
\end{equation}
For rank-2 perturbation, $A$ is the diagonal matrix of radiometer noise contribution and  $U = \begin{bmatrix}
    &  \\
    \vec{q_{0}} &\vec{q_{1}}\\
    &  
\end{bmatrix}$ and $ V^T =
    \begin{bmatrix}
   & \vec{q_{0}}^T& \\
    &  \vec{q_{1}}^T&
\end{bmatrix}$ are the matrices containing first two eigenvectors obtained from SVD analysis.
The rank-2 model of jitter can also be thought in terms of two independent Gaussian processes each arising from one of the two significant eigenvectors $q_{0}$ and $q_{1}$. And the the final jitter noise signal can be expressed as 
\begin{equation}
    \rm
    \vec{n}_{jitter} = \mathbf{U_{q_{0}}}\vec{j_{q_{0}}} + \mathbf{U_{q_{1}}}\vec{j_{q_{1}}} ,
    \label{eq:jitter_GP_rank2}
\end{equation}
where $\mathbf{U_{q_{0}}}$ and $\mathbf{U_{q_{1}}}$ are $(N_{\rm toa} \times N_{\rm epoch})$ matrices constructed out of eigenvector $\vec{q_{0}}$ and $\vec{q_{1}}$ with their corresponding jitter vectors in $\vec{j_{q_{0}}}$, $\vec{j_{q_{1}}}$. Like the rank-1 case, both jitter vectors can be parameterised by $\lambda_{0}$ and $\lambda_{1}$. Use of both the eigenvectors in the model accounts for the decorrelation in the data. Similarly, as the eigenvectors themselves have a frequency dependence, using them also models the frequency dependence of jitter noise.  

\begin{figure*}
 \begin{subfigure}[t]{0.49\linewidth}
    \centering\includegraphics[width=1\linewidth]{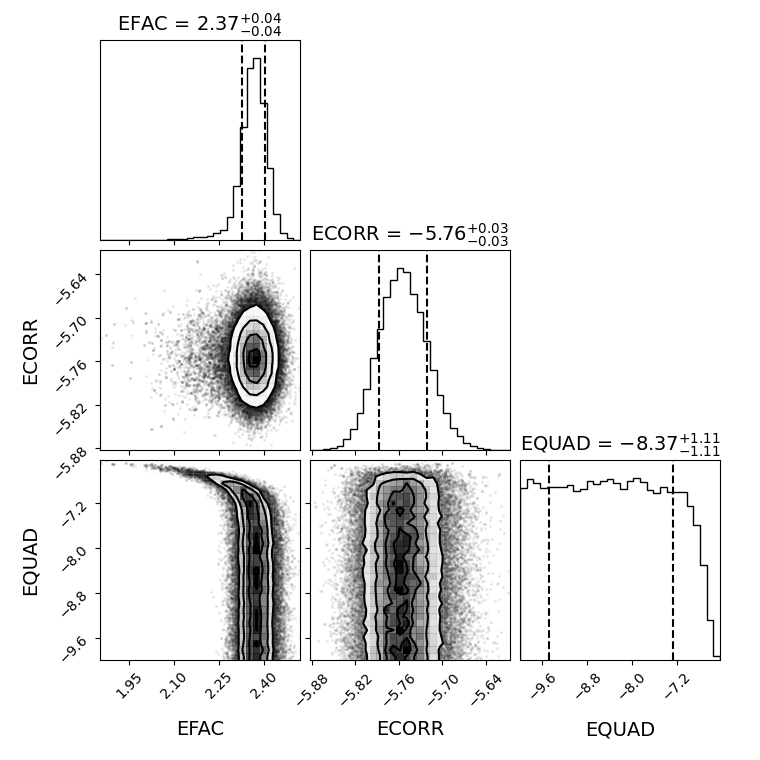}
    
    \caption{rank-1 model}
    \label{fig:4minMK_No_Eig_corner}
     \end{subfigure}
     \begin{subfigure}[t]{0.49\linewidth}
     \centering\includegraphics[width=1\linewidth]{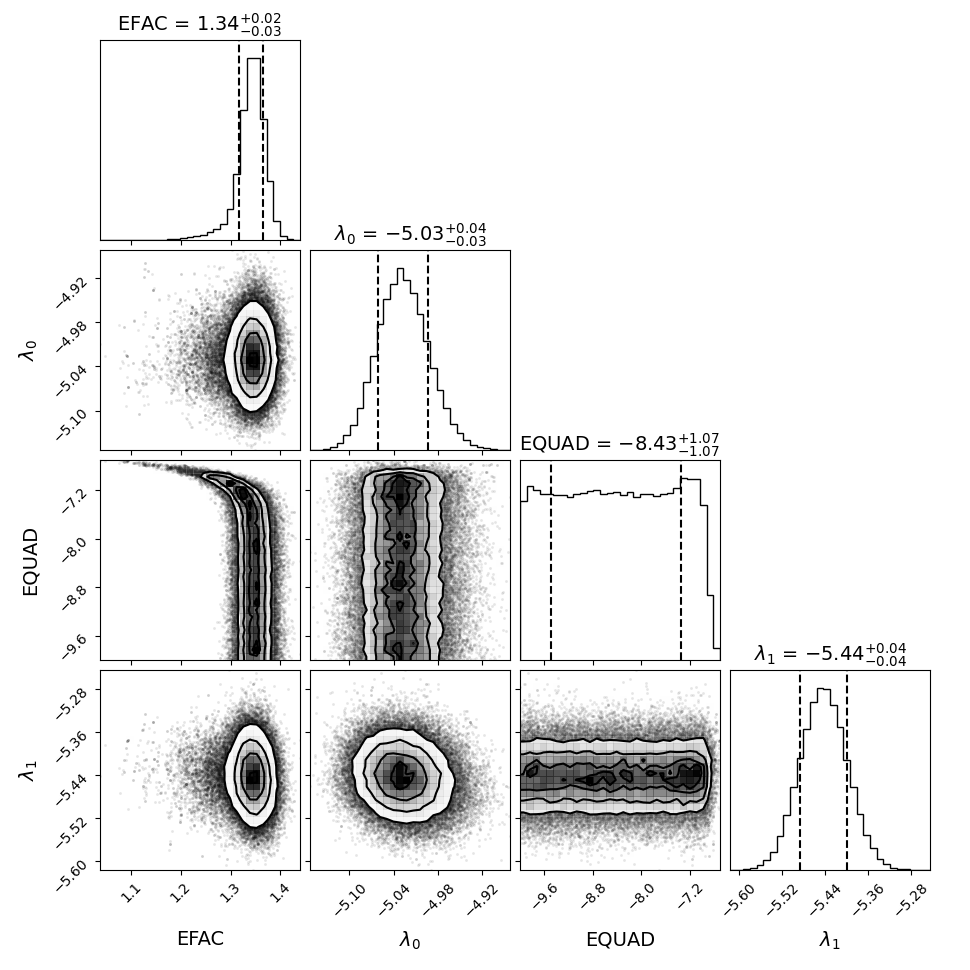}
     
    \caption{rank-2 model}
     \label{fig:4minMK_Eig_corner}
     \end{subfigure}
     \caption{\textbf{MeerKAT 4 min data}: Posterior distribution of white noise parameters. Parameters $\lambda_{0}$ and $\lambda_{1}$ together define the jitter noise.}
\end{figure*}

We then sampled the parameters $\lambda_{0}$ and $\lambda_{1}$ along with other model parameters. 
We initially tested the modified jitter implementation on the same data set which was used to obtain SVD. We selected the prior distribution for the parameters $\lambda_{0}$ and $\lambda_{1}$ to be uniform in logarithmic amplitude. As these data have 4 minute duration, we searched only for white noise parameters. We checked the consistency of results by first sampling the parameter space using \textsc{PTMCMCSampler} \citep{justin_ellis_2017_1037579} to run Markov-Chain Monte-Carlo (MCMC) and also using \textsc{Bilby} \citep{2019ApJS..241...27A} which used a nested sampler `Dynesty' to measure the posterior distribution of model parameters. We compared models by calculating Bayes factor from the evidence values reported by \textsc{Bilby} for each model. Fig~\ref{fig:4minMK_No_Eig_corner} and~\ref{fig:4minMK_Eig_corner} compare the posteriors obtained by using original rank-1 model and with the new rank-2 model. The value of parameter EFAC is significantly lower in rank-2 case. Moreover the rank-2 model is favoured over the rank-1 model with a large log Bayes factor of $ln\mathscr{B} = 1064$. This demonstrates that the rank-2 model better describes the white noise in J0437$-$4715.

We then tested the new jitter implementation with a MeerKAT 4 year timing data of J0437$-$4715. For this test, we searched for the chromatic red noise arising due to long term DM variations. Because of turbulence in the interstellar medium between Earth and the pulsar, the total electron content along the line of sight varies significantly over a timescale of a few years \citep{2007MNRAS.378..493Y}. This causes perturbation in the TOAs which can be modelled as a Gaussian process. The signal due to DM variations can be expressed as 
\begin{equation}
    \vec{n}_{DM} = \mathbf{F_{DM}}\vec{a}_{DM} ,
    \label{eq:DM_GP}
\end{equation}
where $\mathbf{F_{DM}}$ is the Fourier basis comprising sinusoidal terms and $\vec{a}_{DM}$ are corresponding coefficients. As the spatial electron density variations in the ISM are expected to follow the Kolmogorov turbulence, the DM noise can be expressed as a power-law in fluctuation frequency  \citep{2021arXiv210513270T},
\begin{equation}
    P(f) = \frac{{A}^2_{DM}}{12\pi^2} (f/yr^{-1})^{-\gamma_{DM}} yr^3,
    \label{DM_noise_Power-law}
\end{equation}
where ${A}^2_{DM}$ is the amplitude defining the strength of the signal, and $\gamma_{DM}$ is the spectral index which is predicted to be equal to $11/3$. Usually during GWB searches, the coefficients $\vec{a}_{DM}$ are marginalised and only ${A}^2_{DM}$ and $\gamma_{DM}$ are sampled \citep{2014PhRvD..90j4012V}. Another important stochastic signal in pulsar timing is the `spin noise', which is thought to arise due to irregularities in the spin period of pulsars \citep{2010ApJ...725.1607S}. Similar to DM noise, PTAs model spin noise as a Gaussian process with a power-law in fluctuation frequency domain. In contrast to DM noise, spin noise is frequency independent. We performed our tests with and without spin noise in the overall noise model. 
 
\begin{figure*}
    \begin{subfigure}[t]{.49\linewidth}
        \centering\includegraphics[width=1\linewidth]{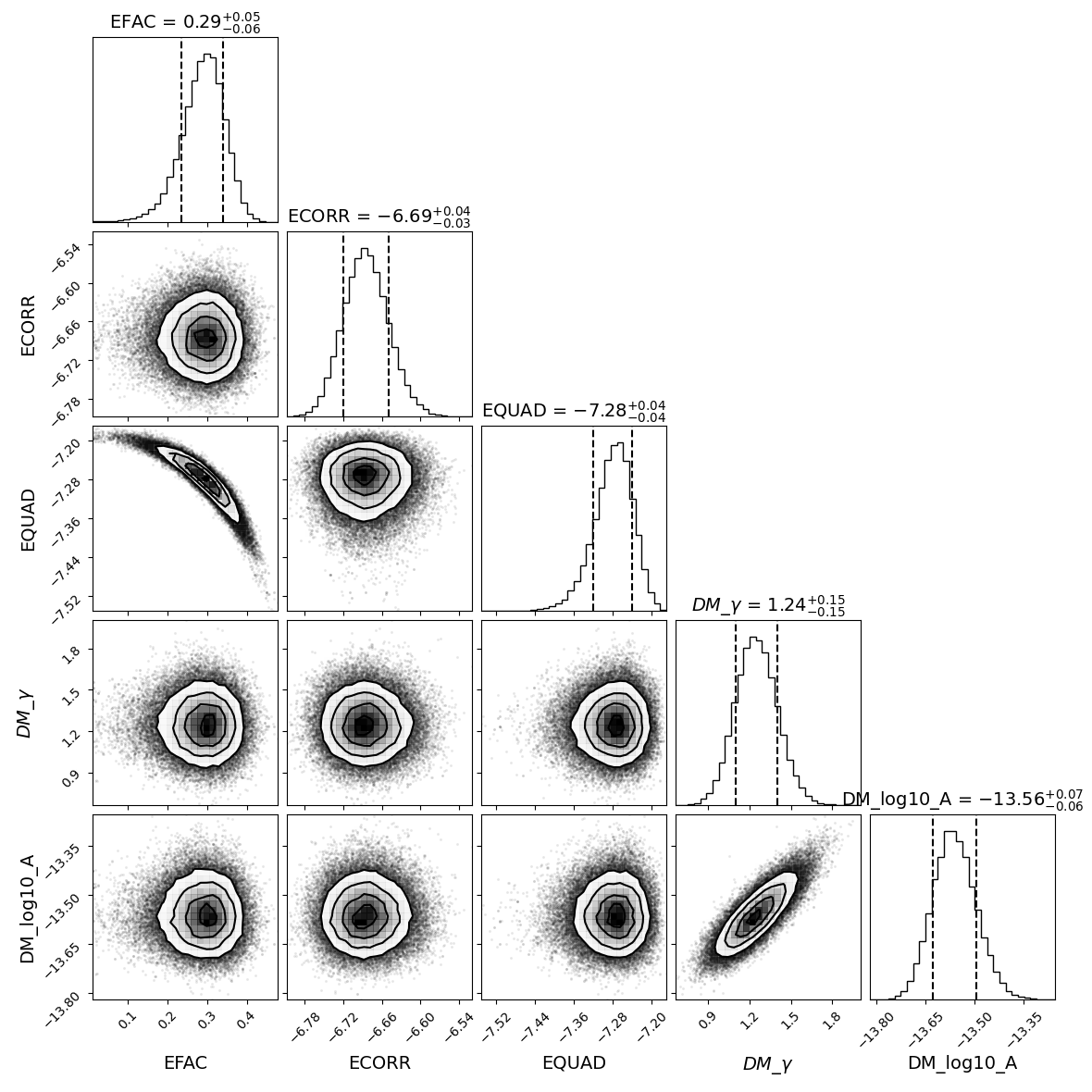} 
        \caption{rank-1 model}
    \label{fig:4yrMK_No_Eig_corner}
    \end{subfigure}
    \begin{subfigure}[t]{.49\linewidth}
        \centering\includegraphics[width=1\linewidth]{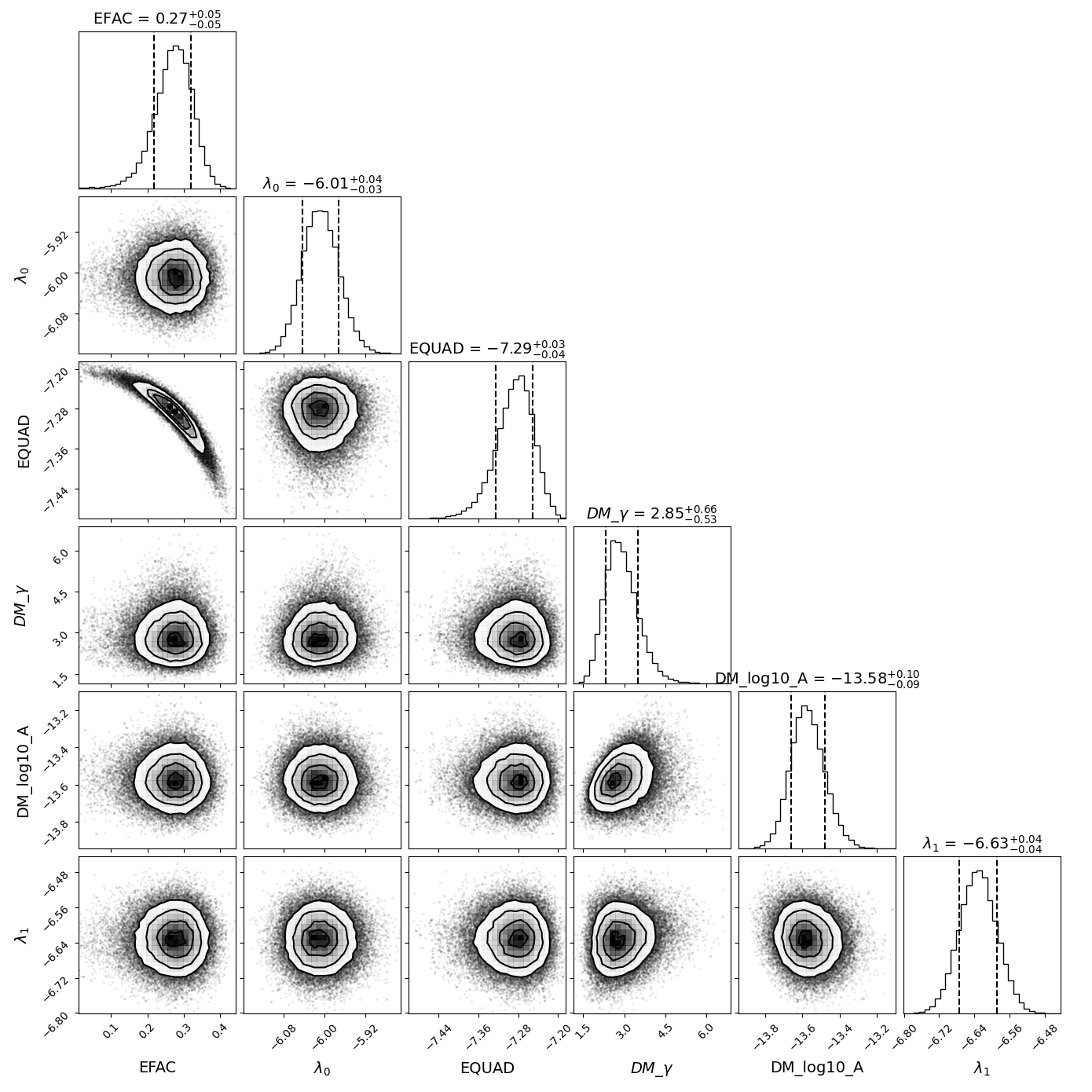}
        \caption{rank-2 model}
    \label{fig:4yrMK_Eig_corner}
    \end{subfigure}
    \caption{\textbf{MPTA 4 year data}: Posterior distribution of white noise and DM noise parameters}
    \label{fig:4yrMK}
\end{figure*}

\begin{figure}
	\includegraphics[width=\columnwidth]{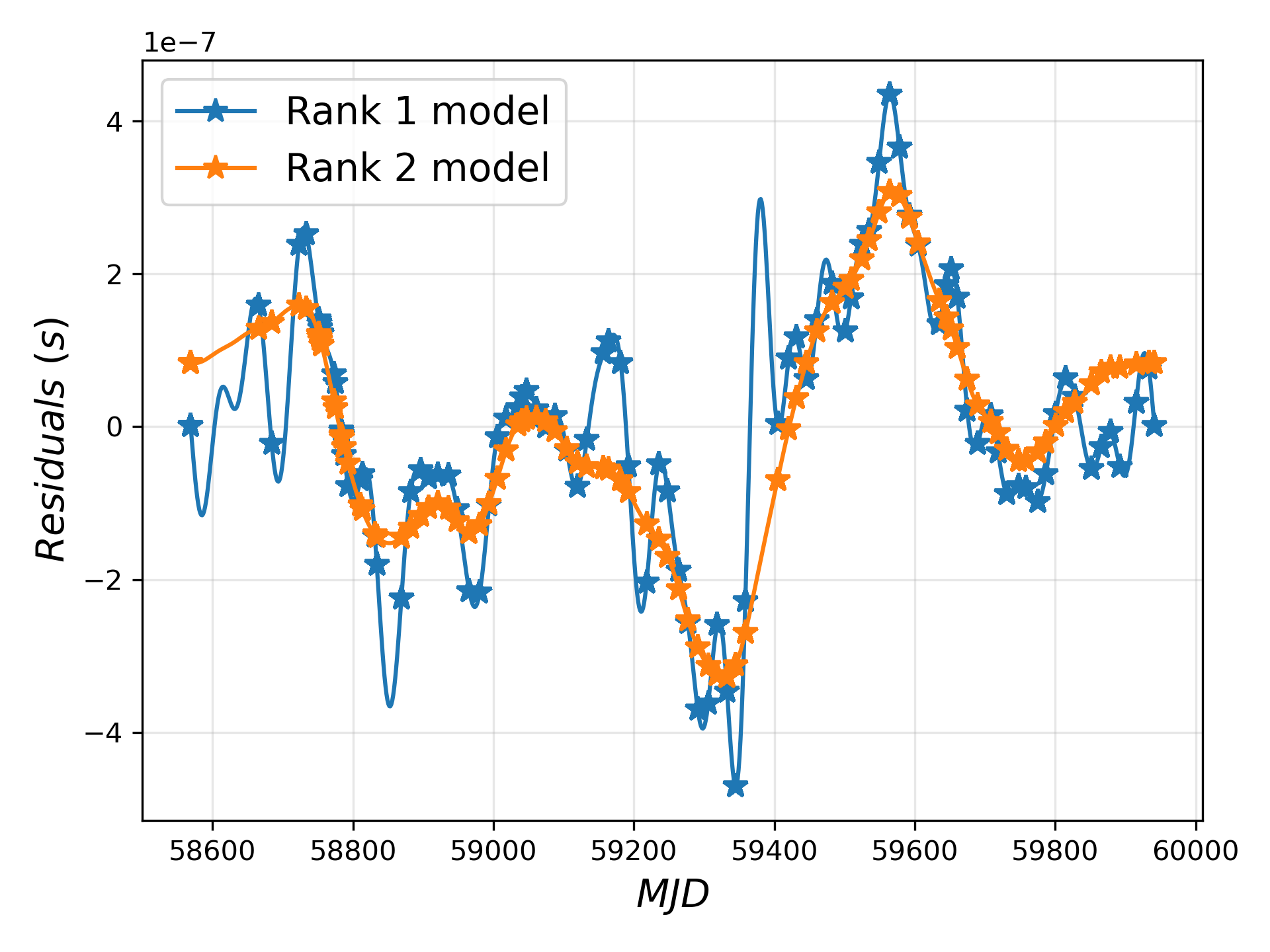}
    \caption{Post fit DM noise realisation with rank-1 and rank-2 model. The points marked with a star denote epochs at which observations were available. The short time scale variations seen in rank-1 case are getting absorbed in the new frequency dependent jitter parameters $\lambda_{0}$ and $\lambda_{1}$.}
    \label{fig:DM_noise_comapre}
\end{figure}

 The main result from these tests is shown in Fig~\ref{fig:4yrMK}. It can be seen that the use of rank-1 model for jitter gives a shallower spectral index for the DM noise. In contrast 
 the rank-2 model results in a measurement of $\gamma_{DM}$ that is more consistent with the expected value. Further, we found that the rank-2 model is favoured over the rank-1 model with a log Bayes factor of $ln\mathscr{B} = 40$. We saw no significant evidence to support the presence of spin noise for this pulsar in our data set .
 It can be argued that since the rank-1 model fails to capture any radio-frequency dependence, the DM noise term is forced to absorb the small timescale radio-frequency dependence in the data which makes the spectral index shallower. But the use of rank-2 model which contains radio-frequency dependent eigenvectors $\vec{q_{0}}$ and $\vec{q_{1}}$, revives the higher spectral index for DM noise. Our result re-emphasises the problem of model mis-specifications in pulsar timing. The inference on deterministic model parameters of a pulsar is known to be sensitive to the values of noise parameters used in the least-squares linear-fit. To see if there were any differences the pulsar ephemeris, we used the Gaussian process method, as described in equation~\ref{eq:jitter_GP_rank2}, to incorporate the rank-2 jitter implementation in the pulsar timing software \textsc{pint} \citep{2021ApJ...911...45L}. We then performed a generalised least-squares fit for astrometric and binary parameters. Simultaneously the post fit values of the DM noise coefficients $\vec{a}_{DM}$were obtained and were used to interpolate the DM variation between the epochs of observation. Results of the fit showed no significant difference in the values or uncertainties of astrometric and binary parameters with the rank-2 model, however the time realisation of the DM noise process was found to be smoother than the rank-1 case. This follows directly from the steeper spectral index for the DM noise in rank-2 model. Fig~\ref{fig:DM_noise_comapre} shows the maximum likelihood DM noise realisation. This suggests that for the rank-1 case, the leakage of frequency dependent short-term fluctuations into the DM noise term would result in incorrect interpolations for the DM variations.

\subsection{Frequency dependence of jitter noise}
\label{subsec:Freq_dependence_jitter}
 The presence of more than one significant eigenvector in J0437$-$4715 is detectable mainly because of the high flux density of this pulsar and the use of sensitive telescopes like MeerKAT. 
 PSR~J0437$-$4715 could be an exception in the wider pulsar population or it could be a first indication of previously unknown physical processes in pulsar magnetosphere which gives rise to a decorrelation of jitter noise. The answer to this question will become clear in the future as we move towards the use of more sensitive telescope receivers with increased collecting area. However, for any other pulsar, even if a single significant eigenvector remains, it can have notable frequency dependence as seen in eigenvector $q_{1}$. Using the NANOGrav 12.5 year data set,  \citet{2019ApJ...872..193L} have performed a detailed study of the frequency dependence of jitter noise in their sample and have shown that 30 out of 48 pulsars are better described by a frequency dependent description of jitter noise. However it is obvious that the use of vector $\vec{u}$ in our jitter model will not be able to capture any frequency dependence. This mis-specification in the model can potentially bias our inference of other chromatic terms like DM variation or the chromatic noise arising due to scattering in ISM, which in turn can affect the inference of gravitational waves in a combined search. Thus it is important that a more complete model for jitter noise should be obtained for the next generation PTAs.

\section{Conclusion}
In this work, we have performed a detailed analysis of spectral de-correlation of the jitter noise as observed in PSR~J0437$-$4715. We show that the decorrelation mechanism is self-similar and the rate of decorrelation can be described using a power-law which depends on the ratio of the frequencies of observation. By performing spectral analysis on the jitter covariance matrix we show that the presence of decorrelation implies that a higher rank-approximation of the jitter covariance matrix is essential in pulsar timing applications. Further if a pulsar shows an appreciable frequency dependence of jitter noise then the current jitter noise model in gravitational wave searches is inadequate to account for it. It is also shown that failing to account for this frequency dependence biases the inference of other frequency dependent terms in the model. The radio emission mechanism in pulsars is a long standing problem in astronomy. The phenomenon of jitter can be used as a probe to the magnetosphere of neutron stars. The analysis, as presented in this work, on the future sensitive measurements of jitter noise in other pulsars can help build more precise models of pulsar radio emission mechanism.

\section*{Acknowledgements}
AK, DR, MM, MS and MB acknowledge support through ARC grant
CE170100004. RMS acknowledges support through ARC Future Fellowship FT190100155. The MeerKAT telescope is operated by the
South African Radio Astronomy Observatory (SARAO), which is a
facility of the National Research Foundation, an agency of the Department of Science and Innovation. PTUSE was developed with
support from the Australian SKA Office and Swinburne University of Technology, with financial contributions from the MeerTime
collaboration members. This work used the OzSTAR national facility at Swinburne University of Technology. OzSTAR is funded by
Swinburne University of Technology and the National Collaborative
Research Infrastructure Strategy (NCRIS)

The Parkes radio telescope is part of the Australia Telescope National Facility 
which is funded by the Australian Government for operation as a National Facility managed by CSIRO. We acknowledge the Wiradjuri people as the Traditional Owners of the Observatory site.

\section*{Data Availability}

The data used for this analysis will be provided on request



\bibliographystyle{mnras}
\bibliography{example} 








\bsp	
\label{lastpage}
\end{document}